\documentstyle[aps,epsfig,twocolumn]{revtex}

\begin{document}

\title{What drives the translocation of stiff chains?}
\author{Roya Zandi$^{1}$, David Reguera$^{1}$, Joseph Rudnick$^{2}$,
and William M.  Gelbart$^{1}$}

\address{$^{1}$Department of Chemistry and Biochemistry, UCLA, Box
951547, Los Angeles, CA 90095-1569\\
$^{2}$Department of Physics, UCLA, Box 951547, Los Angeles, CA
90095-1547}


\maketitle

\begin{abstract}

We study the dynamics of the passage of a stiff chain through a pore
into a cell containing particles that bind reversibly to it.  Using
Brownian Molecular Dynamics simulations we investigate the
mean-first-passage time as a function of the length of the chain
inside, for different concentrations of binding particles.  As a
consequence of the interactions with these particles, the chain
experiences a net force along its length whose calculated value from
the simulations accounts for the velocity at which it enters the cell.
This force can in turn be obtained from the solution of a generalized
diffusion equation incorporating an effective Langmuir adsorption free
energy for the chain plus binding particles.  These results suggest a
role of binding particles in the translocation process which is in
general quite different from that of a Brownian ratchet.  Furthermore,
non-equilibrium effects contribute significantly to the dynamics,
\emph{e.g.}, the chain often enters the cell faster than particle
binding can be saturated, resulting in a force several times smaller
than the equilibrium value.
\end{abstract}

\section{Introduction}

The transfer of genetic material through the membrane surrounding a
cell nucleus is fundamental to the understanding of basic cell
processes from gene therapy to viral infection.  The motion of linear
polymers through pores into confined volumes also arises in many other
biological contexts \cite{Alberts}, perhaps the most common examples of which
include the translocation of proteins from the cytosol into the
endoplasmic reticulum, or into mitochondria or chloroplasts.  The
export of m-RNA through nuclear pore complexes is still another
example of great importance.  Despite the longstanding and widespread
interest in this process, however, our knowledge about it is still
rudimentary.

The process of translocation under influence of an external field or a
chemical potential gradient has recently been studied extensively
\cite{luben,muthu,chern,kardar}.  There have also been several
theoretical studies which have specifically investigated chain
translocation in the presence of binding particles
\cite{oster,park,elston,rap}.  The major role of these binding
particles has been recognized as a ''Brownian ratchet'', a mechanism
which was introduced over ten years ago in pioneering work by Simon,
Peskin and Oster \cite{oster}.  They happen to have treated the case
of protein translocation, but their arguments apply equally well to
nucleic acids.  To account for translocation rates fast compared to
simple diffusion under a wide variety of conditions and circumstances,
they proposed that non-specific binding by globular proteins results
in a \char`\"{}biased\char`\"{} --- or \char`\"{}ratcheted\char`\"{}
--- motion of the chain.  Many experiments confirm that efficient
translocation can indeed take place without the involvement of motor
proteins.  For example, the entire length (about 40 microns) of the
DNA which comprises the genome of T5 phage is observed to enter its
bacterial cell host without requiring metabolic energy \cite{mal}.
The experiment of Salman \emph{et al.} on phage \( \lambda \) DNA
shows similarly that translocation of a comparable length of chain can
occur without the help of active processes; simple diffusion would
require significantly longer times \cite{salman}.

Each of the mechanisms mentioned above for chain translocation, namely
diffusion and ratcheting, corresponds to a different time scale and to
different physics.  Simple diffusion requires a characteristic time \(
t_{d}=L^{2}/2D \), where \( L \) is the total length of the polymer
and \( D \) its diffusion coefficient.  In the ''ratcheting''
scenario, as soon as a specific length --- \( \delta \) --- of the
polymer enters, a protein binds to it and the chain is no longer able
to diffuse backwards because the pore size is too small for the
DNA/protein complex to pass through.  In this case, the chain simply
diffuses from one binding site to the next, and the translocation time
is equal to the product of the time it takes for the chain to diffuse
the distance \( \delta \) times the number of ratcheting sites \(
M=L/\delta \), \emph{i.e.}, \( t_{ratchet}=M\delta
^{2}/2D=L^{2}/(2MD)=t_{d}/M \), corresponding to a speed-up of the
translocation time by a factor of \( M \) over simple diffusion.  As
pointed out by Simon, Peskin and Oster, this time represents an
idealized limit in which the ratcheting mechanism functions
\char`\"{}perfectly\char`\"{}, i.e., as each successive binding site
passes into the cell it is bound irreversibly by a protein that
prohibits the chain from diffusing backwards.  In actuality, however,
an entering site is not necessarily bound immediately by a protein,
and/or the protein does not stay adsorbed long enough to act as a
ratchet at that site; accordingly, the translocation time is increased
beyond \( L\delta /2D \) by a factor that depends on the ratio of on-
and off- rates for binding.

In this paper we consider explicitly the effect of binding particles
and show that --- via a new mechanism --- translocation can occur at
rates significantly faster than that provided by the ''perfect''
ratcheting scenario described above.  More specifically, we argue that
the particles which bind reversibly to the chain give rise to a net
force on the chain that pulls it into the cell.  Furthermore, this
force accounts fully for the translocation process and embraces the
different mechanisms mentioned above, \emph{e.g.}, in a special limit
the Brownian ratcheting appears as a particular idealization (and
appealing simplification) of the effect of such a force.  The
magnitude of the force depends in a delicate way on the concentration
of the binding particles and on their diffusion coefficient relative
to that of the chain.  In the overdamped limit the translocation time
in the presence of a force \( F \) is expected to take the form \(
t_{F}=L/v=L\zeta /F \), where \( \zeta \) is the friction coefficient
of the chain, related to its diffusion constant through the Einstein
relation \( D=k_{B}T/\zeta \).

Using Brownian Molecular Dynamics (BMD) simulations, we calculate
separately the average force and the translocation time in the
presence of binding particles and find that they do indeed obey the
relation \( t_{F}=L\zeta /F \).  Consequently, the translocation
process is force-driven and the translocation time turns out to be
longer or shorter than the ideal ratcheting time, depending on the
concentration and diffusion coefficient of the binding particles.
These results can be understood in the context of a generalized
Fokker-Planck equation for the probability that at time \( t \) the
chain will have entered the cell to a distance \( x \) and have \( n
\) particles bound to it.  The drift terms corresponding to chain
entry and particle binding are obtained from derivatives with respect
to entry distance and binding number of a Langmuir adsorption free
energy for the overall system. \footnote{In the simplest case of the fully
 equilibrated,
fixed chemical potential Langmuir adsorption problem,
 the 1D pressure of the system corresponds to the force
  on the chain: $P_{1D}=f=(k_{B}T/\delta) \ln(1+\phi \exp(\beta \epsilon))$ 
  where $\phi$ is the volume fraction of binding particles, $\epsilon$ 
  their binding energy, and $\delta$ the binding site size.  In the limit 
  of large binding energies $(\beta \epsilon \gg 1)$ one recovers the 
  $f \sim \epsilon / \delta $ result discussed in Section IV; otherwise,
   entropic effects are important as well.  We thank Pierre-Gilles de 
   Gennes for these observations.} 
  We find that the binding process
involves important non-equilibrium effects in general, on which
depends the actual value of the force pulling the chain in, and hence
the ratio \( t_{F}/t_{ratchet} \).

\section{Simulation}

Brownian Molecular Dynamics simulations of a stiff polymer
translocating through a pore and into a cell are performed using a
coarse-grained model in which the chain is represented by a rigid rod
of beads.  The beads are rigidly linked to their nearest neighbors
along the chain and do not interact with each other.  In this way we
model DNA as a perfectly straight --- rather than the usual
semi-flexible --- chain, because the focus of our work is on the
entering segment of chain which is within a persistence length from
the pore.  The link between adjacent beads is rigid in order to avoid
chain contraction and extension resulting from the binding of
particles; we also neglect changes in the shape or twist of the chain
due to binding.  The distance between monomers along the chain, \(
\sigma \), corresponds to the {}``footprint'' of binding particles in
that the center of each bead is considered as an absorbing site.  The
binding particles are modeled as spherical, interacting with each
other through the repulsive part of a Lennard Jones (LJ) potential
with diameter \( \sigma \); the interaction between binding particles
and chain monomers is treated by a \emph{full} (12-6) LJ potential.
Since the distance between the absorbing sites is equal to the
diameter of the binding particles, each site can be surrounded by a
maximum of six particles.  The cell-particle interactions are taken to
vanish for particles within the radius \( R_{s} \) of cell and to
increase as \( (R-R_{s})^{4} \) for particles at distances greater
than \( R_{s} \) from the sphere center.  This potential is simply a
convenient way to describe an interior wall.  Finally, we treat the
pore itself as being completely ''inert'', having no effect on the
chain except to allow it to enter or leave the cell.

\begin{figure}
\centerline{\epsfig{file=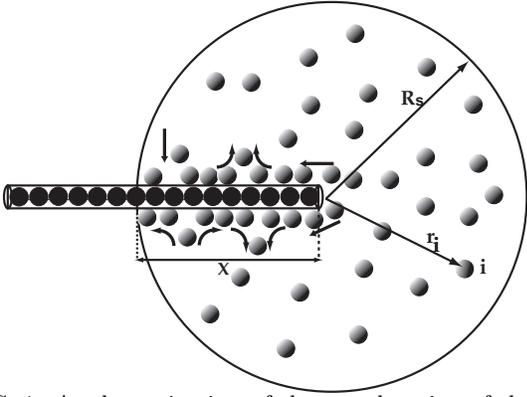,width=7cm}}

\caption{A schematic view of the translocation of the rigid chain
into the
spherical cell.}

\label{fig:cell}
\end{figure}

We focus on the dynamics of the chain once one end has been inserted.
Let \( x \) denote the length of chain inside the cell and \( {\bf
{{r_{i}}}} \) the position of the \( i \) th binding particle (see
Fig.  \ref{fig:cell}).  The time evolution of these coordinates is
described by the overdamped Langevin equations\footnote{We ignore the force in the \( y \) and \( z \) directions on
the chain, and let the rod move only in the \( x \) direction.  }

\begin{eqnarray}
\frac{d{\bf {r}}_{i}}{dt} & = & {\bf {f}}_{i}D_{0}/(k_{B}T)+{\bf {{b}}}_{i}
\label{langevin} \\
\frac{dx}{dt} & = & FD_{rod}/(k_{B}T)+B \, \, .
\end{eqnarray}
Here \( {\bf {{f_{i}}}} \) and \( F \) are the deterministic forces
acting on each particle and the rod, respectively, and \( {\bf
{{b_{i}}}} \) and \( B \) are the corresponding random ---
{}``Brownian'' --- forces satisfying \( <{\bf {{b_{i}}}}(t)>=0 \), \(
<{\bf {{b_{i}}}}(t)\cdot {\bf {{b_{j}}}}(t')>=6D_{0}\delta
(t-t^{\prime })\delta _{ij} \), \( <B(t)>=0 \) and \(
<B(t)B(t')>=2D_{rod}\delta (t-t^{\prime })> \).  \( D_{0} \) is the
diffusion coefficient of an individual binding particle, related to
its friction coefficient, \( \zeta _{0} \), through the Einstein
relation \( D_{0}=k_{B}T/\zeta _{0} \).  As for the chain, we
introduce an effective diffusion coefficient, \( D_{rod}=k_{B}T/\zeta
_{rod} \) which in principle may include all the pore-DNA
interactions.  Since so little is known about these complicated
interactions, we have simply taken \( D_{rod}=D_{0}/(L/\sigma ) \),
consistent with the translational diffusion coefficient of a stiff
chain being inversely proportional to its length.  In all that follows
we use \( \sigma \), \( k_{B}T \), and \( \sigma ^{2}/D_{0} \) as the
units of length, energy, and time.  \( \epsilon \), the LJ binding
energy between the particles and the monomers, is set equal to \(
5k_{B}T \); the diameter of the spherical cell is \( 2R_{s}=24\sigma
\) and the total length of the chain is \( L=16\sigma \).

Figure \ref{fig:cell} is a schematic snapshot of the simulated system.
The particles bind predominantly at the tip and then move back along
the chain, with particles occasionally adding in empty spaces along
the rod; at the same time, particles leave from other parts of the
chain and allow for new particles to bind.  Geometrically, up to six
particles can sit around each monomer, but just as in the familiar
Langmuir adsorption problem there are always empty sites on the chain
due to entropic factors.  The fact that particles are mostly added at
the tip is completely a dynamical effect.  Under the influence of
particle binding, the chain is in general moving too fast to allow for
saturation of adsorption along its length.  This results in less than
the equilibrium number of occupied sites on the chain.  In general,
particles bind to the tip of the chain by pushing aside some already
attached particles.  However, the force pulling in the chain is 
exerted mostly at the entering positions where empty binding sites first
appear.

\section{Results}

\begin{figure}[t,b]
\centerline{\epsfig{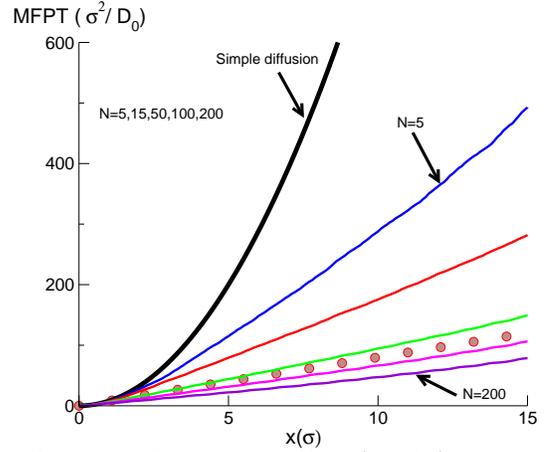}}

\caption{Mean First Passage Time (MFPT) as a function of entry
distance \protect\( x\protect \), for each of several different values
of the number of binding particles \protect\( N\protect \), as
calculated from our BMD simulations.  The thick solid curve describes
the MFPT \emph{vs} \protect\( x\protect \) for simple diffusion of the
chain into the cell.  The dotted curve shows the MFPT for the case
where translocation would occur via perfect ratcheting (see text).}

\label{fig:timedens}
\end{figure}

Figure \ref{fig:timedens} shows the Mean First Passage Time (MFPT)
versus the length of the chain inside the cell, \( x \).  MFPT is the
average, over a large number of trajectories, of the time it takes
for the front tip of the chain to first arrive to the position \( x
\).  Each curve in the figure corresponds to a different value of \( N
\), the number of binding particles inside the cell.

As mentioned in the INTRODUCTION, there exist different mechanisms for
the translocation of the rod under the influence of the binding
particles.  If the chain simply diffused into the spherical cell, the
MFPT would be equal to \( t_{d}=L^{2}/(2D_{rod})=2048\sigma ^{2}/D_{0}
\) corresponding to the length \( L=16\sigma \); see the quadratic
function depicted by the heavy solid curve in Fig.
\ref{fig:timedens}.  As shown in the figure, \( t_{d} \) lies
significantly above the translocation times we find in our
simulations; even in the presence of only five binding particles, the
MFPT is about three times shorter.

The dotted line in Fig.  \ref{fig:timedens} represents the time that
it would take for the entire chain to enter if there were ratchets
functioning perfectly at every absorbing site.  In this case, the
chain simply undergoes successive and independent diffusions between
neighboring sites, completing each in a time \( \sigma ^{2}/(2D_{rod})
\); the MFPT is equal to the product of this time and the number of
steps \( (x/\sigma ) \) associated with the entry distance \( x \).
According to the ratcheting mechanism, then, the slope of time versus
displacement is simply \( \sigma /(2D_{rod}) \) =$8 \sigma /D_{0}$, for
$D_{rod}=\frac{D_{0}}{16}$.

The numerical results from our simulations, as well as the theory
outlined in the next section, confirm the presence of a quite
different translocation mechanism, namely, drift due to a net force
exerted on the chain by binding particles.  Entry into the cell of
successively longer portions of chain {}``feeds'' new binding sites
into the system; as each additional particle binds to the chain the
free energy of the chain drops, and this reduction gives rise to a
force pulling the chain into the cell.  In the presence of a constant
force, and in the overdamped limit, the corresponding translocation
time is \( t_{F}=L/v \) where \( v=FD_{rod}/k_{B}T \) is the velocity
of the chain.

With competing mechanisms operative, the MFPT will reflect
predominantly the one with the smallest translocation time.  As
already remarked, the diffusion time is always much longer than those
arising from the other two mechanisms (see Fig.  \ref{fig:timedens}).
However, the ratcheting time could be longer or shorter than that of
the force-driven process according to whether the deterministic force
\( F \) is larger or smaller than the {}``effective'' Brownian
ratcheting force.  Comparison of the ratcheting velocity of a chain,
\( 2D_{rod}/\sigma \), with the usual expression for the velocity of a
chain under a constant force, \( v=F_{ratchet}D_{rod}/k_{B}T \), shows
that the {}''effective'' Brownian ratcheting force is \( 2k_{B}T/\sigma
\).  Consequently, the ratcheting mechanism is expected to be dominant
in the presence of weak enough driving forces.  Nevertheless, even in
this limit, as we shall see below, the attractive force is operative
and completely determines the translocation velocity.

Several important observations can be extracted from Fig.
\ref{fig:timedens}.  The first is that translocation times depend on
the number (concentration) of binding particles.  This is also true,
of course, for the ratcheting mechanism, since higher concentrations
of binding particles imply faster {}``on-rates'' (\( k_{+} \)), hence
higher translocation velocities.  But, as Simon, Peskin and Oster have
themselves emphasized, there is a maximum translocation rate
corresponding to the limit of large \( k_{+}/k_{-} \).  This maximum
rate --- or minimum time, \( t_{ratchet}=L\delta /2D_{rod} \) --- also
corresponds to the smallest distance (\( \sigma \) in the present
model) between ratcheting (binding) sites.  From Fig.
\ref{fig:timedens} we see that the translocation occurs even faster
than the limiting ratcheting prediction when the number of binding
particles exceeds \( N=100 \).  This result suggests that an
additional mechanism is operative, which we show is associated with a
net force acting on the chain along its direction of motion.

More explicitly, the slope of MFPT \emph{vs} \( x \) plots reveals
that the average velocity of the chain, \( v \), remains almost
constant throughout the translocation process (except right at the
beginning and towards the end).  From the relation \( F=\zeta _{rod}v
\) one expects that the average force on the chain also stays constant
during this process, in which case the slope of time \( t \) versus \(
<x> \) will be almost the same as that of MFPT versus \( x \) iff \(
vx/D_{rod}\gg 1 \)%
\footnote{The translocation time of a polymer can be defined in terms
of the mean first passage time \( \tau (x) \), which satisfies the
backward Fokker-Planck equation with a reflecting boundary condition
at the hole and an absorbing boundary condition at \( x \).  The
solution to the backward Fokker-Planck equation in the presence of a
constant force \( F=\zeta v \) is \( \tau
(x)=\frac{x}{v}+\frac{D}{v^{2}}(e^{-vx/D}-1) \).  If \( v/D\gg 1 \),
then \( \tau (x)=\frac{x}{v} \).  In the other limiting case,
\emph{i.e.}, when \( v/D\ll 1 \), then \( \tau (x)=\frac{x^{2}}{2D} \).
}.  Using \( F=\zeta _{rod}v \), we calculated the effective force on
the chain from the velocity obtained in our MFPT \emph{vs} length
curves (Fig.  \ref{fig:timedens}).  Specifically, we calculated the
average velocity of the chain over the range \( x=4\sigma \) to \(
x=11\sigma \), \emph{i.e.}, in the region where the MFPT versus \( x
\) curves are almost linear (see Fig.  \ref{fig:timedens}).

\begin{figure}
\centerline{\epsfig{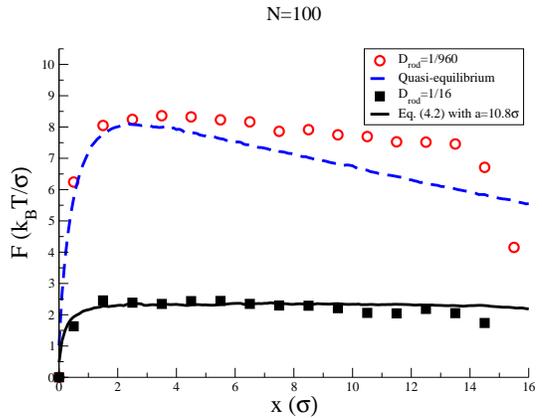}}

\caption{The squares illustrate the force calculated directly in the
simulation (coarse-grained over\protect\( \sigma \protect \)), as a
function of the length of the chain inside, for the case \protect\(
N=100\protect \).  The circles show the force calculated from an
identical BMD simulation but with a rod diffusion coefficient which is
16 times smaller.  The dashed curve corresponds to the force derived
from solution of the coupled entry/adsorption diffusion equation in
the approximation of quasi-equilibrium.  The solid curve is the
solution of the full Fokker-Planck Eq.  (\ref{2D Fokker-Planck}) using
\( a=10.8\sigma \) (see text).}
\label{fig:force60}
\end{figure}

\begin{figure}
\centerline{\epsfig{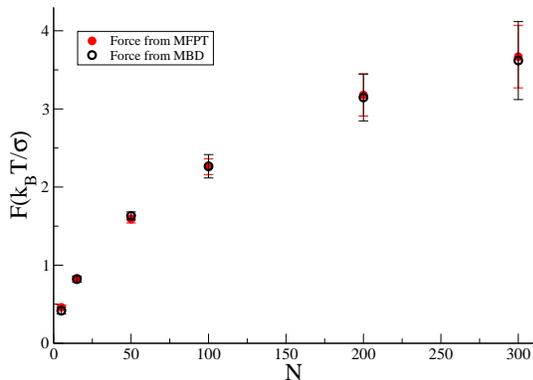}}

\caption{Comparison of the force obtained directly in the simulation
(hollow circles) with the one calculated using the relation \protect\(
F=\zeta _{rod}v\protect \) (filled circles) where \protect\( v\protect
\) is the velocity corresponding to the inverse slope of MFPT
\emph{vs.  \protect\( x\protect \)} data.\label{fig:compare}}
\end{figure}

Alternatively, we can also determine the force acting on the chain
directly from the simulations.  The squares in Fig.  \ref{fig:force60}
represent the average force exerted on the chain by the binding
particles (coarse-grained over \( \sigma \)), as a function of the
length inside.  This force is seen to be nearly constant; furthermore,
it is found to agree within numerical uncertainty, as shown in Fig.
\ref{fig:compare}, with the force calculated (as described above) from
the average velocity (\emph{i.e.}, the inverse slope in
Fig.\ref{fig:timedens}) for all the values of \( N \) (up to 300) that
we treated.

For \( N<50 \), where the forces are of the order of unity or smaller
(i.e., smaller than the {}``effective'' Brownian ratchet force \(
2k_{B}T/\sigma \)), Fig.  \ref{fig:timedens} shows that the
translocation times begin to be significantly smaller than the perfect
ratcheting limit.  This scenario is in principle embodied in the basic
result of Ref.  \cite{oster} in which the translocation time is
written as \( t_{ratchet}(1+2K^{-1}) \) where \( t_{ratchet}=L\delta
/2D_{rod} \) is the ideal ratcheting time and \( K=k_{+}/k_{-} \) is
the effective strength of binding, expressed as a ratio of ''on'' and
``off'' rates.  \( K \gg 1 \) corresponds to saturated binding and to
ideal ratcheting; otherwise, the ratchet mechanism becomes less
efficient and the translocation times are longer than \( t_{ratchet}
\).  It is notable that even in this weak force situation, where one
expects the ratcheting mechanism to dominate, we still observe a
force-controlled translocation (\emph{i.e.,} the translocation time is
still determined by the force pulling the chain in, as shown in Fig.
\ref{fig:compare}).  The reason is that these weaker forces correspond
to smaller numbers of less strongly bound particles, and hence also to
less efficient ratcheting.

\section{Dynamical Theory}

The translocation process simulated above can be described
theoretically using a simplified dynamical model.  The two relevant
variables are the length of the chain \( x \) inside the cell and the
number of particles \( n \) attached to the chain.  As the chain
enters a length \( x \), the number of available absorbing sites
increases.  With \( \epsilon \) the binding energy of a single
particle we can write the (Langmuir adsorption) free energy, \( A \),
of the system as

\begin{eqnarray}
\label{energy}
A(x,n)=-n\epsilon -k_{B}T\log \frac{(6x/\sigma )!}{n!(6x/\sigma
-n)!} \\ \nonumber 
-k_{B}T(N-n)\log \frac{V}{(N-n)v_{0}}\, \, .
\end{eqnarray}

Here \( N \) is the total number of particles, \( V \) is the volume
of the spherical cell, and \( v_{0} \) is the volume of a single
particle.  The coefficient 6 in the second term of Eq.  (\ref{energy})
is the number of particles that can interact attractively with a
single binding site (chain monomer); accordingly, \( 6x/\sigma \) is
the total number of available sites on a chain of length \( x \).  The
first term in Eq.  (\ref{energy}) is the energy gain due to binding;
the second is the entropic contribution associated with the total
number of ways in which \( 6x/\sigma \) sites can be occupied by \( n
\) particles; and the last term is the (ideal gas) contribution
associated with the {}``free'' particles, numbering \( N-n \).

We consider the dynamics of translocation as a coupled diffusion
process involving both the \( x \) and \( n \) degrees of freedom.
Using Mesoscopic Nonequilibrium Thermodynamics (MNET) \cite{MNET}, one
can derive the Fokker-Planck equation governing the time-dependent
probability density \( \rho (x,n,t) \) that at time \( t \) a length
\( x \) of the chain has passed through the hole and has \( n \)
particles attached to it:

\begin{eqnarray}
\label{2D Fokker-Planck}
\frac{\partial \rho (x,n,t)}{\partial t}=\frac{\partial }{\partial
x}D_{rod}\left( \frac{1}{k_{B}T}\frac{\partial A(x,n)}{\partial
x}\rho +\frac{\partial \rho }{\partial x}\right) \\ \nonumber
+\frac{\partial
}{\partial n}D_{n}\left( \frac{1}{k_{B}T}\frac{\partial
A(x,n)}{\partial n}\rho +\frac{\partial \rho }{\partial n}\right) .
\end{eqnarray}

\noindent Note that there is a drift and a diffusion term for each of
the \( x \) and \( n \) variables.  The force driving the
translocation is \( F(x,n)=-\frac{\partial A(x,n)}{\partial x} \), and
we see from Eq.  (\ref{energy}) that its origin is entropic, arising
from the second ({}``Langmuir'') term in Eq.  (\ref{energy}); the
binding of the particles gives rise to a force that pulls the chain
in.  Similarly, the factor \( -\frac{\partial A(x,n)}{\partial n} \) is a
{}``thermodynamic force'' driving the particle binding.  \( D_{rod} \)
is the spatial diffusion coefficient of the rod (with the usual
dimensions of \( length^{2}time^{-1} \)), while \( D_{n} \) (which in
general may depend on \( x \) and \( n \)) is the \emph{kinetic} rate
constant (with dimensions of \( time^{-1} \)) for the process of
particle binding and unbinding.  A crude, but time-honored and
physically reasonable, expression for \( D_{n} \) comes from the
Smoluchowski theory of aggregation dynamics \cite{chandra} for
diffusing particles:

\begin{equation}
\label{D_{n} aproximate}
D_{n}=a\frac{N}{V_{cell}}D_{0},
\end{equation}
  where \( a \) is a length of order the particle size. \( D_{n} \)
here is simply proportional to the concentration of binding particles
\( N/V \), and their spatial diffusion coefficient \( D_{0} \).

The Fokker-Planck Eq.  (\ref{2D Fokker-Planck}) provides a complete
description of the kinetics of both chain entry and particle binding.
However, a simpler description can be achieved by considering the
possibility of time scale separation.  The characteristic times for
the entry and the binding processes scale as \( \tau _{x}\sim
1/D_{rod} \) and \( \tau _{n}\sim 1/D_{n} \), respectively.  If the
binding process is very fast compared to the chain entry (i.e.  \(
\tau _{n}/\tau _{x}\sim D_{rod}/D_{n}\ll 1 \)), it is reasonable to
assume that the {}``fast'' variable --- here the number of attached
particles, \( n \) --- will decay very rapidly to its equilibrium
distribution.  In this case the process can be described by the
evolution of the slow variable, the position \( x \) of the chain.
Suppression of the fast variable can be carried out using the standard
adiabatic elimination technique \cite{risken}, which is essentially
equivalent to integration of the Fokker-Planck equation over the
equilibrium distribution of the fast variable.  The resulting one
dimensional Fokker-Planck equation is

\begin{equation}
\label{1D Fokker-Planck}
\frac{\partial \widetilde{\rho }(x,t)}{\partial t}=\frac{\partial
}{\partial x}D_{rod}\left(
\frac{\widetilde{F}(x)}{k_{B}T}\widetilde{\rho }+\frac{\partial
\widetilde{\rho }}{\partial x}\right) ,
\end{equation}
  where

\begin{equation}
\label{drift}
\widetilde{F}(x)=\int \frac{\partial A(x,n)}{\partial x}f_{eq}(n;x)dn,
\end{equation}
  is the average driving force. Here we have assumed that the spatial
diffusion coefficient of the rod is independent of $n$, and defined \( \rho
(n,x,t)=\widetilde{\rho }(x,t)f_{eq}(n;x) \), with \( f_{eq}(n;x) \)
the {}``local'' equilibrium distribution,\begin{equation}
\label{local equilibrium}
f_{eq}(n;x)=\frac{e^{-\frac{A(x,n)}{k_{B}T}}}{\int
e^{-\frac{A(x,n)}{k_{B}T}}dn}.
\end{equation}

In the particular case where \( f_{eq}(n;x)=\delta (n-n_{eq}(x)) \),
the force driving the translocation process becomes \(
F_{eq}(x)=\left.  \frac{\partial A(x,n)}{\partial x}\right|
_{n_{eq}(x)} \), and the number of attached particles is equal to the
equilibrium one, \( n_{eq}(x) \), given by the solution of \( \left.
\frac{\partial A(x,n)}{\partial n}\right| _{n_{eq}}=0.  \)

The above quasi- (or {}``local'') equilibrium approach basically
assumes that, as soon as the chain's advance makes available new
sites, particles bind to them immediately.  Figure \ref{fig:force60}
compares, for \( N=100 \), the average force calculated in our
simulation (squares) with the {}``quasi-equilibrium'' force given by
Eq.  (\ref{drift}) (dashed curve).  The comparison shows that the
actual force that the binding particles exert on the chain is
significantly smaller than the one which follows from the assumption
that binding equilibration can keep up with chain entry (and this is
true for all other \( N \) values considered).  To check the limit in
which the simple quasi-equilibrium 1D Fokker-Planck description,
(\ref{1D Fokker-Planck}), provides an accurate description of the
translocation process, we slowed down the entry of the rod by
decreasing \( D_{rod} \).  The circles in Fig.  \ref{fig:force60}
shows the average force calculated via BMD simulation when \(
D_{rod}=1/960 \) (\emph{vs} 1/16).  Agreement with the
''quasi-equilibrium'' force evaluated from (\ref{drift}) is now
excellent, confirming the validity of our model for the thermodynamic
free energy \( A(x,n) \).  Comparably good agreement is found between
the number of attached particles calculated in the simulation and that
predicted by the quasi-equilibrium theory.  (Alternatively, we can
increase the diffusion coefficient \( D_{0} \) of the binding
particles, in which case the calculated force also approaches its
quasi-equilibrium value, as we checked in the simulation).  The free
energy given in Eq.  (\ref{energy}) was written in the continuum limit
under the assumption that the system is dilute.  Considering its
simplicity, the accuracy of the model in describing the system is
surprisingly good.

In general, to compare simulation results with the predictions of our
theory, we need to solve the full Fokker-Planck equation for the
coupled \( x \) and \( n \) degrees of freedom.  The numerical
solution is obtained by converting Eq.  (\ref{2D Fokker-Planck}) into
its equivalent set of Langevin equations \cite{risken}, and then
solving these equations using standard stochastic algorithms.  From
these numerical solutions the average force, the mean first passage
time, and the average number of adsorbed monomers can be determined.
Using the approximate expression for \( D_{n} \), given by Eq.
(\ref{D_{n} aproximate}), with \( a=10.8\sigma \) (which in fact is very
close to the value \( a=4\pi \sigma \) from the Smoluchowski theory)
we have solved Eq.  (\ref{2D Fokker-Planck}) for different values of
\( N \).  The resulting MFPT values \emph{vs} \( x \), and averaged
pulling forces, compare well with those from the simulations.  As an
example, the resulting average force for \( N=100 \) is represented by
the solid line in Fig.  \ref{fig:force60}, showing an excellent agreement
with the results of the simulation.

\section{Conclusion}

The present work has attempted to provide a basic theoretical
framework for treating the translocation dynamics of a stiff chain as
it moves into a cell containing particles which interact attractively
with it and bind to it.  We conclude that this process is in general
\emph{force}-controlled.  Obviously, \emph{pure} diffusion occurs only
in the absence of any binding particles; and the \emph{rectification}
of diffusion ---which is the essence of the Brownian ratcheting
mechanism--- appears as a manifestation of the binding force in a
special limit of interaction potentials and particle concentration.

To examine the extent to which the simple Brownian ratchet mechanism
can account for chain translocation due to binding particles, we have
performed several simulations using different values for \( \delta \),
the distance between binding sites.  We varied \( \delta \) from \(
1\sigma \) to \( 4\sigma \), \emph{i.e.}, from \( \delta \) smaller
than to larger than the range of the Lennard-Jones attractive
interaction (about \( 2\sigma \)) .  When \( \delta \) is sufficiently
larger than the range of interaction --- say for \( \delta =4\sigma \)
--- we find that the chain performs a cycle of drift and diffusion
movements.  As an absorbing site enters the cell the chain is first
pulled by the attractive force acting on the binding site, and then
diffuses between adsorbing sites, during which time no net force is
acting on the chain.  In this case, the effect of the binding force is
twofold: it pulls the chain in the region where it is acting, and also
impedes its backwards diffusion.  In the limit where the range of the
force is small compared to the distance between sites, the
contribution of the drift to the translocation is negligible, but
nevertheless the force is still rectifying the diffusion.  In these
circumstances, the effect of the force can be most simply described by
the Brownian ratchet idea.  Specifically, the rectification arises
from the free energy penalty for moving an absorbing site out of the
cell (as provided approximately by our Langmuir adsorption model).
The efficiency of the rectification depends on the ratio of free
energy penalty to the thermal energy \( k_{B}T \).

We have also found that the effective force can be significantly
(several times) smaller than the value one would estimate from a
quasi-equilibrium treatment of the binding and entry dynamics, i.e.,
from assuming that the time scale for particles to diffuse and bind is
much shorter than that for chain entry.  More explicitly, only if we
increase sufficiently the friction coefficient of the chain do we find
agreement between the force calculated from the simulations and that
obtained from the quasi-equilibrium solution to the coupled
entry/binding Fokker-Planck equation.  This result indicates that
there can be significant differences between the \emph{stalling} force
measured in single-molecule experiments and the actual \emph{value} of
the force during the process of translocation.

In summary, as established directly from our BMD simulations, we find
that chain translocation can be understood in terms of the pulling
force arising from particle binding.  Moreover, this mechanism leads
to translocation times that can become distinctly smaller than
ratcheting.  By decreasing the diffusion coefficient of the chain
relative to that of the binding particles, the simulations give a
still larger effective force pulling the chain along its length into
the cell.  The maximum value of this force, attainable in the limit \(
N \gg 1 \) and \( x\rightarrow \infty \) is precisely \( 6\epsilon
/\sigma \), \emph{i.e.} the drop in free energy per unit length when
all binding sites are occupied.  These results are nicely confirmed by
solving directly a coupled Fokker-Planck equation for the chain entry
and particle binding dynamics.  We conclude that physically realistic
situations are in general more complicated than a ratcheting mechanism
in which it is assumed that the only effect of particle binding at a
chain site is to prohibit its diffusing back through the pore.

The authors would like to acknowledge helpful discussions with Y.
Kantor, M. Kardar, P. G. de Gennes, J. Widom and M. Deserno.  This research
was supported by the National Science Foundation under Grants No.
CHE99-88651 and CHE00-76384.

\end{document}